# 3D Medical Image Segmentation based on multi-scale MPU-Net

Zeqiu.Yu[1,2], Shuo.Han[1], Ziheng.Song[3]

[1]Department of Statistics and Data science, Northwestern University, Evanston, IL, 60208, USA
[2]Department of Radiology, University of Pittsburgh, Pittsburgh, PA 15213, USA
[3]School of Engineering and Applied Sciences, Columbia University in the City of New York, New York, NY 10027, USA

## Abstract

The high cure rate of cancer is inextricably linked to physicians' accuracy in diagnosis and treatment, therefore a model that can accomplish high-precision tumor segmentation has become a necessity in many applications of the medical industry. It can effectively lower the rate of misdiagnosis while considerably lessening the burden on clinicians. However, fully automated target organ segmentation is problematic due to the irregular stereo structure of 3D volume organs. As a basic model for this class of real applications, U-Net excels. It can learn certain global and local features, but still lacks the capacity to grasp spatial long-range relationships and contextual information at multiple scales. This paper proposes a tumor segmentation model MPU-Net for patient volume CT images, which is inspired by Transformer with a global attention mechanism. By combining image serialization with the Position Attention Module, the model attempts to comprehend deeper contextual dependencies and accomplish precise positioning. Each layer of the decoder is also equipped with a multi-scale module and a cross-attention mechanism. The capability of feature extraction and integration at different levels has been enhanced, and the hybrid loss function developed in this study can better exploit high-resolution characteristic information. Moreover, the suggested architecture is tested and evaluated on the Liver Tumor Segmentation Challenge 2017 (LiTS 2017) dataset. Compared with the benchmark model U-Net, MPU-Net shows excellent segmentation results. The dice, accuracy, precision, specificity, IOU, and MCC metrics for the best model segmentation results are 92.17%, 99.08%, 91.91%, 99.52%, 85.91%, and 91.74%, respectively. Outstanding indicators in various aspects illustrate the exceptional performance of this framework in automatic medical image segmentation.

# Contents

# 1 Introduction

Malignant tumors have long posed a severe hazard to human health and life. In the present globalization trend, terminal cancer has been one of the study foci of researchers from numerous nations as the disease with the greatest mortality rate. The worldwide burden of cancer can be quantified as follows: In the preceding year, the number of new cancer diagnoses was close to 20 million, and approximately 10 million people died of cancer, of which advanced cancer accounted for a significant part [1]. With the economic expansion, countries are actively improving cancer care technology. Furthermore, early identification and treatment of precancerous cancer can significantly enhance a patient's survival percentage [2], and the method for determining the tumor's location has therefore become one of the most widely utilized procedures in medical research. Typically, this technique is referred to as medical image segmentation since it is capable of accurately determining the tumor's location and plays an indispensable part in tumor radiotherapy and treatment [3]. The research of this technology has substantially increased the efficiency and accuracy of tumor outlining, specifically:
1. Reduce the strain on doctors by preventing them from manually outlining hundreds of serial slices with high subjectivity.
2. To avoid misdiagnosis as a result of clinicians separating healthy tissues into lesion regions.
3. Explicitly present the patient's treatment effect.

Medical image semantic segmentation is commonly categorized into three categories in regular clinical practice as a primary way of targeting and treatment: manual, semi-automatic, and automatic segmentation [23]. For the first two, clinical researchers need to distinguish between distinct organs and tissues in the human body. This kind of physical intervention is restricted not only by the expert's experience, but it is also easy to make mistakes while completing such a huge workload. This sort of segmentation technology, on the other hand, has limited feature representation capabilities, and it's difficult for an image segmentation architecture designed for one type of organ to perform better in another. As a result, automatic medical image semantic segmentation is favored and assumes a dominating position in practical applications in order to prevent error-prone extensive labeling of medical images, enhance the work efficiency of clinical researchers, and minimize their burden.

In today's era of the rapid growth of computer-aided treatment and diagnosis, the segmentation technique of tumor images is increasingly inclined to full automation. This technology has supplied clinicians with a wealth of analytical resources for preoperative research and treatment plans [3, 4]. Image recognition and classification [5, 6], machine translation [7, 8], target identification [9, 10], sentiment analysis [11, 12], and artistic creation [13, 14] have all achieved notable success with the assistance of deep learning. Segmentation technology of images is a significant study path in computer vision science, as well as an important component of image semantic understanding. The remarkable ability of CNN is reflected in the non-linear feature extraction, which makes it a leading choice in this sector. Although a neural network that solely relies on CNN has a lot of redundant calculations between patches, the architecture based on it has a significant impact on medical image processing [15, 21, 22]. Image segmentation technology evolved gradually after the fully convolutional neural network (FCNN) was introduced [16]. The image segmentation's technological application in the medical industry is still quite complex due to the diverse tumor forms, individual variances, vast dispersion, and sensitivity to noise during imaging. Medical picture segmentation, nevertheless, has far-reaching implications in clinical research as the initial stage in assessing anatomical structure. The computer replicates the human visual perception mechanism and uses the computer's strong computational capability to swiftly produce correct analysis and processing findings [5], allowing for faster and more effective clinical communication.

Among the numerous variants of FCNN, the U-Net based framework has demonstrated exceptional representation ability in the testing of diverse medical image datasets [17, 18, 19, 20], and has been steadily and extensively employed in the medical image segmentation field. For instance, by redefining dense connection, Dolz et al. [22] presented HyperDense-Net for multi-modal brain tissue segmentation. Zhou et al. [17] proposed an encoder-decoder sub-networks composed of a stacked of stacked dense hop path connections, for CT scan image liver segmentation and polyp segmentation. Strong Two-Pathway-Residual blocks were suggested by Aghalari et al. [18] for automated segmentation of MRI images of brain malignancies. These clinical investigations' segmentation tasks are usually divided into two steps: positioning and segmentation [24]. The image obtained by patient positioning is irreversible. Therefore, the image with high positional accuracy has a great positive effect on segmentation, which will improve cancer physicians' diagnosis and treatment of patients. It is evident that the existing segmentation technology for medical images is not evolved enough to assist clinical practices comprehensively, and this is a field that many researchers are working to explore and improve.

Although this type of FCN-based network design produces advanced results, the convolutional layer's intrinsic locality restricts the network's capacity to replicate long-distance correlation coupling. Therefore, this sort of architecture performs poorly in multi-scale information structure segmentation and is unable to capture the non-local backdrop well. The establishment of the self-attention method of the convolutional layer can effectively boost the global modeling skills [21, 27], but Transformer's excellent capabilities for sequence-to-sequence prediction make it a new substitute. The Transformer is based entirely on the attention mechanism, eliminating loops and convolutions [28], and has exhibited outstanding performance in machine translation and medical image semantic segmentation [8, 29]. Nevertheless, even though it is capable of learning long sequence information, the secondary time complexity and intrinsic restrictions of the codec structure prevent it from being directly used for Long Sequence Time-series Forecasting (LSTF) [30].

In 2019, Fu et al. [92] developed the Dual Attention architecture, in which the spatial dimension decreases quadratic computational cost while simultaneously allowing for effective capture of global feature interdependence. To create spatial dependencies between any two locations, learning the correlation of spatial properties is frequently accompanied by sophisticated attentional connection structures. Through suitable weight assignment, the Position Attention Module in Dual Attention realizes the mutual correlation of identical characteristics in any two positions, and this relationship is no longer related to the distance between them. When dealing with complicated scenarios, our flexible and lightweight architecture can capture the semantic properties of inconspicuous items and differentiate related traits.

In this paper, we attempt to be the first to propose a combination of PAM architecture and multi-scale attention mechanism to investigate the possibility of semantic segmentation in medical images. For this purpose, this work puts forward the MPU-Net, a novel framework for volumetric medical image semantic segmentation. Based on reformulating the challenge of volume segmentation as a long sequence prediction problem in one-dimensional space with the use of the U-Net framework, the attention block will be utilized as a part of the encoder to grasp the global relationship of the context from the embedded input block. To put it simply, the suggested framework in this study re-establishes the self-attention mechanism from the standpoint of sequence. By skipping the connection and connecting directly to the decoder, the combination of elements with different resolutions will be accomplished to realize the precise location. Alongside the capacity of employing PAM to accomplish global information collection, a decoder based on multi attention block will be employed to obtain comprehensive spatial resources at different resolutions to fill the gaps in feature resolutions information produced by the encoder.

The following are the primary contributions of this research:
1. To enhance the effectiveness of medical image semantic segmentation, this research offers a novel 3D medical image model based on the U-Net framework. The overview of MPU-Net is demonstrated in Figure 1.
2. In this research, a Position Attention Module is utilized as a method for effectively capturing long-distance dependent information in volumetric medical pictures.
3. The U-Net nested network has been rebuilt. After sampling the PAM's self-attention features, a skip-connection decoder is utilized to fuse distinct high-resolution CNN features to produce the predictive segmentation output.
4. The primary structure of FCNN, the fast convolutional layer, is replaced with a Hierarchical Convolutional Neural Network (HCNN) to increase the efficiency of retrieving long-term sequence features.
5. A novel loss function is suggested in this study, which is a mix of Tversky loss and cross-entropy. With the MPU-Net framework, the loss of the model is still in a downward trend after convergence.
6. To improve the network, a cross-attention mechanism based on multi-scale blocks is introduced after each upsampling, and multi-scale output is employed to accomplish the fusion of feature information from various resolutions.
7. Test the effectiveness of the suggested framework on the challenge of volumetric medical image segmentation using the public dataset: Liver Tumor Segmentation Challenge 2017 (LiTS 2017). The comparison to the benchmark model UNet architecture demonstrates our model's advantage in segmentation.

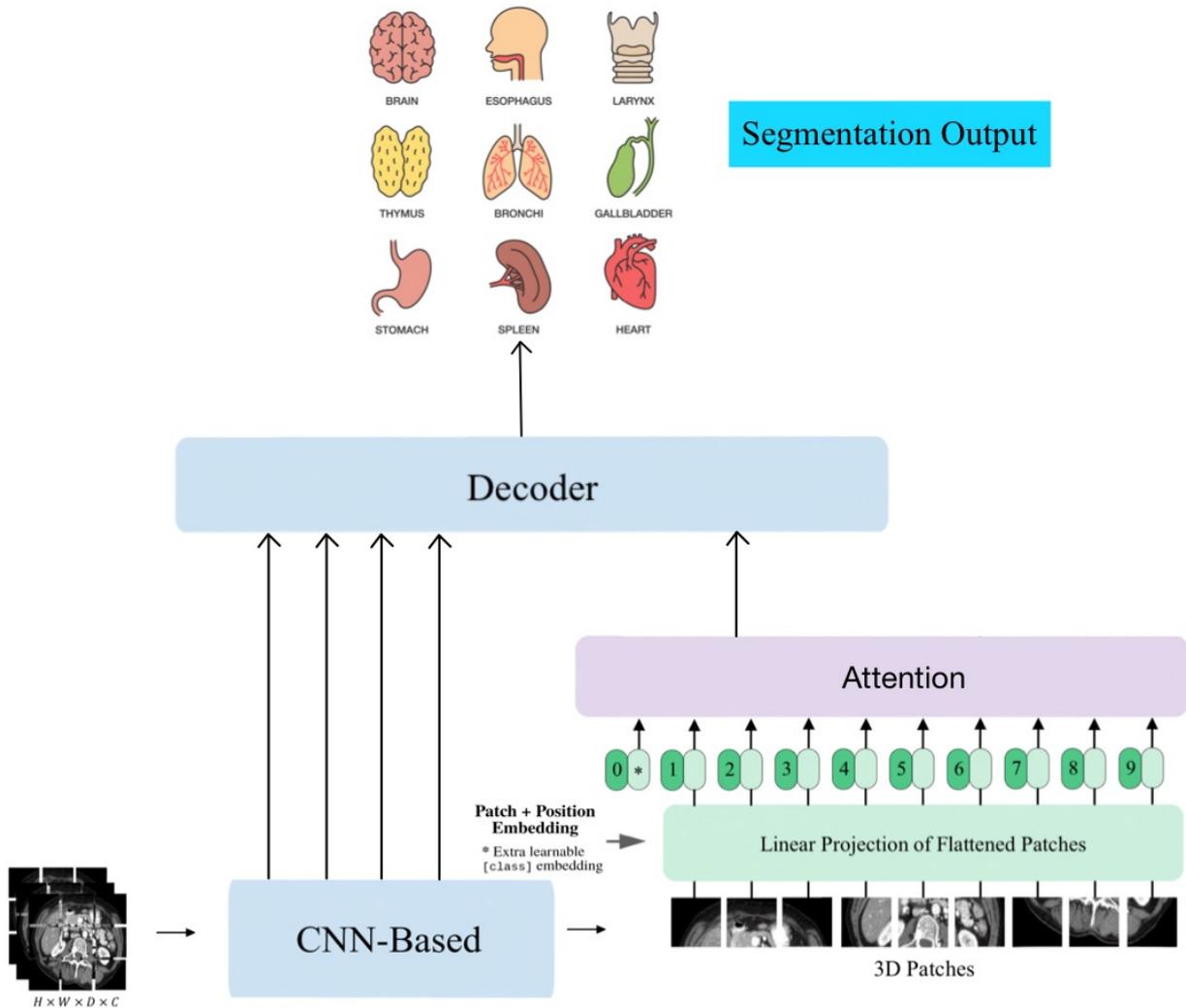

Figure 1: The rough architecture of the MPU-Net framework presented in this study.

The key sections of this article are grouped as follows: The next portion will go through the scientific research achievements that have been completed. The suggested segmentation approach for the MPU-Net architecture will be detailed in the third section. The experimental environment background in which the model is tested, as well as the various evaluation measures for the model, are introduced in section 4. Section 5 delves deeper into the findings of the experiment, including quantitative analysis and visualization. Sections 6 and 7 will combine theory and experiment for a more in-depth discussion and conclusion. The code for performing the experiment can be found in the appendix.

# 2 Literature Review

## 2.1 Image Semantic Segmentation

Image segmentation technologies have always received extensive attention from scholars. As a kind of basic computer vision technology, image segmentation separates the image into different sections with varied features from which the items of interest are extracted. It is a crucial component of image analysis and the foundation of image understanding. Various image segmentation approaches have been devised to apply to various sorts of patterns throughout the previous century [33]. Traditional image segmentation techniques may be classified into two groups: (1): Operate based on the information of the current target. Such methods can make use of relevant prior information, but they do not rely on prior knowledge. Many conventional methods fall into this category, including edge detection [34, 35], region-based [36, 37], and segmentation techniques based on the combination of edges and regions [38, 39]. This approach can only extract the image's local characteristics, and it's difficult to incorporate the image's prior knowledge into the understanding mechanism of high-level information [40]. (2): Model-based image segmentation technologies directly rely on prior knowledge. This strategy is more in tune with the human visual system [41]. This sort of technique, in particular, may integrate the low-level visual features of the segmented image into high-level features, and then complete the segmentation task by modeling specified targets. Before deep learning made a great influence on image segmentation algorithms, the more traditional segmentation algorithms were based on statistical models. For example, the Chan–Vese model, Piecewise Smooth model, and Local Binary Fitting model have demonstrated promising outcomes in image semantic segmentation [42, 43, 44]. Over time, neural networks based on the CNN architecture have steadily taken over the area of image processing since the launch of AlexNet [45]. Deep learning has progressed with the times since then, promoting the development of image processing. Image segmentation technology has ushered in huge innovation [16, 51], accompanied by a series of extremely top image analysis works [6, 9, 46-50] in various sectors such as image classification and identification.

In recent years, image detection utilizing deep learning neural networks has grown in popularity as a result of the rapid development of technologies such as image recognition and target identification. Deep CNN has begun to employ selective search to construct candidate areas to determine the position of the bounding box since Girshick et al. [52] introduced an R-CNN architecture integrating AlexNet and SVM. Specifically, images may be seen via windows of various sizes, and adjacent pixels can be clustered together to identify things of interest based on texture, color, or intensity. Although this selective search method substantially accelerates the detection of targets, it necessitates the training of three independent models: a convolutional neural network for generating image characteristics, a regression model for constructing narrower bounding boxes, and a classifier for predicting target categories. This makes training the model extremely challenging. Girshick [53] made two enhancements the next year and presented the Fast R-CNN model: (1): Apply Region of Interest Pooling (RoIPool) to allow candidate regions to share CNN findings, eliminating thousands of CNN calculations through maximum pooling. (2): Combine the above three models into one system. Particularly, to input the bounding box coordinates, the SVM classifier is applied to replace the linear regression layer on top of the original collector, which is parallel to the Softmax layer. In this way, all of the essential inputs are provided by one network, which outperforms R-CNN in practical applications [54]. However, the process of applying selective search to build candidate areas in FastR-CNN is still quite sluggish. Ren et al. [55] employed a convolutional feature layer to produce candidate areas at a low cost, then used Fast R-CNN to construct and categorize smaller bounding boxes. In the realm of image segmentation, this novel candidate area network instantly showed superior performance in the competition [56]. Scholars have since begun to investigate new technologies for locating the pixels of each target, that is, image segmentation. The breakthrough was the redefinition of RoIPool resulting in the proposal of Mask R-CNN, which combines the acquired mask with the original classifier and regressor to accomplish the precise segmentation job. This technique has yielded excellent results in image segmentation applications in different fields [58, 59].

Although image segmentation algorithms have been continuously innovated over the last several years, there are few deep learning algorithms especially built for medical image semantic segmentation. In summary, there exist several difficulties: (1): Medical datasets are difficult to gather, particularly annotated samples that might indicate negative or positive results [60]. (2): Medical image segmentation always focuses on a specific organ or region, making it challenging for a segmentation algorithm to get decent results when segmenting many tissues. At the same time, determining the border between the target organ and surrounding tissues may be difficult. (3): The image acquisition in the positioning stage prior to the medical image segmentation job of clinical illness detection is susceptible to noise and the moving image of the patient's internal organs, causing the image to be easily blurred or uneven [24]. (4) During the training of the model, issues such as over-fitting, a long training duration, and gradient disappearance may arise. It's still challenging to train a deep neural network with strong generalization capabilities [61].

Medical pictures are therefore difficult to segment effectively due to these qualities,

making them exceedingly unstable in practical practice. Many methods concentrating on medical image segmentation have been presented to tackle these problems. Such as threshold technique, region growing, region splitting and merging, classifier and clustering, random field method, and statistics approach [62]. These approaches are all for manual or semi-automated segmentation, and they lack generic feature extraction methods and adaptability in their algorithms. As a result, a completely automated segmentation system for medical image semantic segmentation is urgently needed. In 2015, the Fully Convolutional Network (FCN) was presented to be capable of performing pixel-level prediction challenges using end-to-end training [16]. Soon the FCN-based model showed superior capacity in various medical image segmentation works [60,63]. Simultaneously, it has steadily gained a foothold in the semantic segmentation of medical images.

## 2.2 U-Net Architecture

In 2015, Ronneberger [89] headed a team that developed the U-Net framework. As a substantial variant of FCN, it is currently one of the most prominent strategies in the realm of medical image semantic segmentation. Although this model does not have fully connected layers, this mechanism permits the deeper layer to handle the pixel categorization issue the shallower high-resolution layer meets the pixel classification challenge. *ReLU* is followed after repeating two 3x3 convolutions. The feature channel is then halved by using the maximum pool sampling aid. The expansion path includes an upsampling, which is connected to a feature map of the corresponding contraction path. Then there are two convolutions and a *ReLU*. Each element's feature vector is converted to a class label in the final 1x1 convolution. The whole symmetrical architecture's skip link guarantees that U-Net retains all contextual information, and it has proven the best performance in various 2D and 3D medical picture segmentation tasks [17-20, 60, 73, 83, 106].

In clinical manifestations, U-Net has demonstrated superb capabilities. However, physicians cannot determine the characteristics of tumors or target organs just by gray value since some pathologies lack evident traits in 2-dimension. For 3D, it changes depending on the status of blood vessels, and irregularities in 3D form might provide effective assistance in clinical medical diagnosis [64]. To speed up convergence and eliminate network structure bottlenecks, 3D U-Net employs batch normalization based on 2D U-Net, with adjustments performed when the number of channels doubles and the deconvolution procedure begins. This approach, in particular, may produce a full complete picture immediately from 2D slices. 3D U-Net reconstructs the image mask in the decoder after obtaining critical information by reducing the loss function in the encoder. Then in order to fully leverage the 3D method, it is required to specifically use ultra-small or super-large materials to create negative pictures on stage. In short, only local characteristics may be detected if the receptive area is too tiny. However, if the sensation field is too large, an excessive amount of inaccurate data will be received. Isensee et al. [65] devised a multi-scale model architecture to address this issue. This technique was soon incorporated into clinical studies. In the segmentation challenge of liver tumors, Kushnure et al. [66] employed a multi-scale strategy to shrink the complexity of computation and the parameters of the network, which improved the network's segmentation performance by extending the reception field of CNN. Although this sophisticated U-Net variant architecture has shown positive results, the size of the feature maps has been expanded while preserving high-resolution feature maps to boost segmentation speed. This is not conducive to expediting training and lowering the complexity of optimization in the later stage. An attention mechanism thus needs to be introduced to overcome this problem.

## 2.3 Attention Mechanism

Deubel [71] headed the team that proposed the notion of an attention mechanism in the last century, with the goal of getting the system to concentrate on the most crucial elements while ignoring the rest. The attention mechanism was successfully implemented into the RNN model for image categorization by Mnih et al. [72] in 2014, allowing the machine to learn attention. Then in 2018, the attention mechanism in computer vision was initially postulated by Wang et al. [70]. Since then, this structure has been frequently employed in natural language processing, image processing, statistical learning, multi-modal tasks, self-supervised learning, and other fields of core technology [67], when paired with the application of deep learning. The attention mechanism, which replicates human visual information processing systems by paying attention to motions or notable places in the scene, is essential for achieving optimal allocation of information processing resources [68]. During this procedure, static will be momentarily disregarded. From the standpoint of a model, this technique can focus on high-weighted significant information while ignoring low-weighted relatively irrelevant information. The weight can be modified continually throughout the operation, resulting in increased scalability and resilience. This mechanism is commonly employed as a component to promote the overall effectiveness of the system by sharing selected vital information to exchange with other information in order to complete the target information transfer [69].

In 2017, the Google machine translation team where Vaswani [28] worked proposed the self-attention mechanism for the first time. It is a variant of the attention mechanism, which aims to tackle the problem of long-distance reliance by assessing the mutual effect between different information. It lessens reliance on external information and enhances the ability to capture internal data or feature correlations when compared to the attention mechanism. As an essential part of Transform, the self-attention mechanism is critical in machine translation tasks [28]. Since then, the self-attention mechanism has become a hot spot among neural network attention researchers, with breakthroughs in voice recognition [74], disease prediction [75], image reconstruction [76], and other areas.

Many researchers have attempted to combine the self-attention mechanism with neural networks over the last two years, which helps to accomplish effective modeling of remote dependency. For example, Zhang et al. [77] obtained long-distance perceptual dependency information through high computing capability by combining the self-attention mechanism and CNN, exhibiting strong framework scalability and the ability to efficiently capture contextual semantic information from the long-term framework in ransomware classification tasks. [78] fused LSTM with the self-attention mechanism, which realized the modeling of global information through the chemical characteristics of small molecules, and showed outstanding performance on multiple data sets. Mi et al. [79] exploited the self-attention idea to compensate for the Generative Adversarial Network's (GAN) incapacity to acquire global information efficiently, which demonstrated great resilience to diverse types of pictures in varied experimental conditions.

The self-attention mechanism has acquired prominence as a result of its success in a variety of disciplines. However, there are few studies on medical image segmentation with the self-attention mechanism. Su et al. [80] created a novel GAN architecture by combining the residual connection and the segmentation network with the self-attention mechanism to accomplish exact prostate segmentation. Valanarasu's [81] team took full advantage of the self-attention mechanism's distant reliance and introduced additional control mechanisms to this module to improve the present model's performance and thoroughly learn global and local properties. Taken together, the medical image segmentation algorithm that incorporates the self-attention mechanism is more sensitive and accurate at label prediction, and to a certain extent shows extraordinary versatility.

## 2.4 Transformer

The Transformer has been advancing in the field of NLP with its unprecedented performance since its usage in machine translation in 2017 [28]. The introduction of Transformer to the work of medical image segmentation considerably enhances the accuracy of medical imaging computer-aided diagnosis. In the segmentation task of 3D brain tumors, Wang et al. [82] et al. employed a novel encoder coupled with Transformer, which not only efficiently collects 3D local context information but also models global features well. The Transunet developed in [73] takes into account the strengths of U-Net and Transformer, which maximizes the performance of transformers designed for sequence-to-sequence, producing unprecedented results in tasks such as multi-organ and heart segmentation. Hatamizadeh et al. [83] use the Transformer to make jump connections across multiple resolutions in 3D tasks and link straight to the decoder, successfully capturing global multi-scale information. In addition, Petit et al. [84] also presented the Self and Cross attention mechanisms to redefine the global interaction between features as well as the method for filtering non-semantic features. The strong performance in abdominal image segmentation demonstrates the attention mechanism and the transformer's significant improvement over the U-Net architecture.

# 3 Methodology

## 3.1 Overall Architecture of MPU-Net

A contraction-expansion model is used in the framework provided in this article. As shown in Figure 3, this architecture is established in the space of spatial resolution image $x \in R^{H \times W \times L \times C}$, where $H$ is the dimension of height, $W$ is the dimension of width, $L$ is the dimension of depth/length, and $C$ is the number of channels. To acquire the spatial and depth information of the volumetric medical image, this work employs a mixture framework composed of CNN and Position Attention Module as the encoder is employed here to forecast the pixel-level image of $H \times W \times D \times C$. The image will be delivered to the decoder, which employs skip connections to decode the spatial resolution after it has been encoded into a high-level feature representation through the feature map.

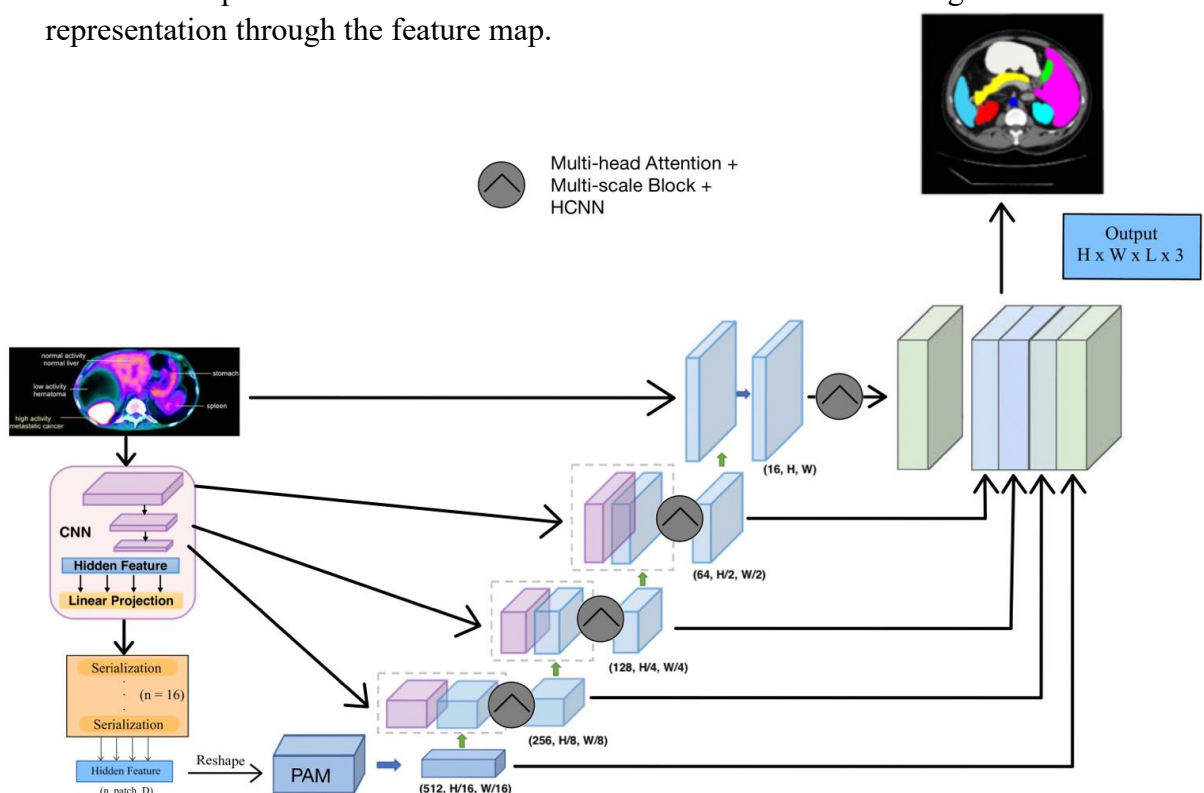

Figure 2: Overview of the proposed multi-scale MPU-Net framework.

On the encoder side, models with long-distance dependence will be suggested, and then a cascaded upsampler will be used for upsampling. High-resolution segmentation results can be achieved when a series of operations are conducted on each enhanced feature layer. Multi-scale prediction is then blended into the overall architecture, and by aggregating the characteristics of multiple intermediate layers, more comprehensive global knowledge of various scales can be obtained. Specifically, to screen and fuse features retrieved at different scales, the HCNN framework and multi-scale blocks are employed. The last layer is the enhanced feature layer, which integrates information from the four-layer network at various resolutions to efficiently reduce noise and promote the final segmentation result.

Next, the various components of the MPU-Net architecture will be elaborated.

## 3.2 Network Encoder

Convolutional Neural Networks (CNN) are introduced into the model. Here, instead of directly employing the Position Attention Module (PAM) as the encoder, a CNN-PAM hybrid model will be used to match the U-Net network architecture to comprehensively improve the positioning accuracy. In particular, Section 3.2.1 will show that $(H \times W \times L) / P^3$ is usually significantly less than the initial image resolution $H \times W \times L$. This will result in the loss of information, demonstrating that utilizing PAM as the encoder is not the ideal option.

In the architecture designed in this paper, CNN is initially applied as an extractor to produce feature maps for the input of PAM. This strategy will help to efficiently employ multiple degrees of resolution in the decoder. To begin, it's necessary to require a one-dimensional input embedding sequence as the input of PAM architecture.

### 3.2.1 Image Sequentialization

This part draws on the experimental analysis and discussion of the TransUNet architecture by Chen et.al [73]. Because the computing complexity of utilizing the Transform framework is quadratic in comparison to the sequence length, reshaping the input 3D medical picture into a flat sequence as the Transformer's input is impracticable. Here, Transformer could be replaced by the PAM architecture, and the approach of partitioning the image into fixed-size patches and reshaping is still highly successful at lowering complexity [85]. Simultaneously, this strategy prevents PAM from constructing a local context information model across places. Based on Vaswani et al. [28] 's idea of image serialization, in this paper, the three-dimensional input information $x \in R^{H \times W \times L \times C}$ will be tokenized by dividing it into a flat and uniform patch sequence

$$X = \{x_p^i \in R^{P^3 \cdot C} \mid i = 1, 2, 3, \cdots\cdots, N\}, \tag{1}$$

where the size of each patch is $P \times P \times P$, and the sequence length (ie: the number of image patches) is $N = (H \times W \times L) / P^3$.

In this way, a one-dimensional sequence with resolution $(H, W, L)$ and $C$ channel numbers can be generated. Here, each slice contains the local volume context feature, and $(P, P, P)$ also represents the resolution of each slice. The PAM encoder will then utilize the whole $X$ of these slices as the input data to further gain more about the global context information and realize the modeling of remote dependencies.

### 3.2.2 Embedding Local Feature Information

Based on the given feature map $X$, the vectorized patch $x_P$ can be projected into the $d$-dimensional space to increase the channel size by using a trainable linear layer (a $3 \times 3 \times 3$ convolutional layer). Moreover, the space and depth dimensions need to be folded into one dimension to generate a $d \times N$ feature map $x$ since PAM only accepts a sequence as input. In order to cipher the location data of each patch, the feature embedding needs to be introduced that can retain the spatial information and the feature map $f$:

$$z_0 = x + PE = W \times X + PE = [x_p^1 E; x_p^2 E; x_p^3 E; \cdots\cdots; x_p^N E] + E_{pos}, \tag{2}$$

where $W$ is a linear projection operator, $E \in R^{(P^3 \cdot C) \times D}$ is the embedding projection of the patch, $PE = E_{pos} \in R^{d \times N}$ represents a one-dimensional learnable position embedding, and $z_i \in R^{d \times N}$ is a specific feature embedding.

After being transformed into a 1D image sequentialization, the original 3D image will be molded into $x_p^i$. The meaning of $N$ is to split the three-dimensional image's length, width, and height by $P$ to get a sequence of patches of size $P^3 \cdot C$. Then, via a linear transformation, they are projected into the space of dimension d. That is, the 3D picture with the original size of $H \times W \times L$ is flattened as the vector $x_p^i$ by the moniker N with the size of $P^3 \cdot C$.

This approach can be likened to a convolution process with kernel size $P \times P \times P$ and step size $P$ on an image of input size $H \times W \times L \times C$. The output size $j$ for input size $i$ according to the convolution output calculation technique, is:

$$j = \left\lceil \frac{i + 2 \ast p - k}{s} + 1 \right\rceil, \tag{3}$$

Here $p$ is the padding, $k$ is the kernel size, and $s$ is the stride. Therefore, if we take

the image's length, width, and height into account, we can get:

$$\left[\frac{H+0-P}{P}+1\right]=\left[\frac{H}{P}\right], \quad \left[\frac{W+0-P}{P}+1\right]=\left[\frac{W}{P}\right], \quad \left[\frac{L+0-P}{P}+1\right]=\left[\frac{L}{P}\right], \quad (4)$$

$N$ is derived from their product, which also represents the equivalent of dividing a 3D image into $N$ patch modules of size $P^3 \cdot C$.

### 3.2.3 Position Attention Module

The PAM Layer is made up of $j-th$ sample layers, each with a unified structure. The output of the Informer layer can be calculated as:

$$x_{j+1}^i = MaxPool\left(ELU\left(Conv\,1d\left([x_j^i]_{AB}\right)\right)\right), \quad (5)$$

where $[*]_{AB}$ contains the basic operation of the previous layer, $Conv\,1d(*)$ is a time-domain one-dimensional convolution filter with a kernel width of 3.

Géron's [86] recommendation is used here to overcome the scenario where the standard activation function $ReLU$ outputs 0 when x is smaller than 0: In the case that the network architecture can prevent self-normalization, the $ELU$ activation function proposed by Clevert et al. [87] can avoid gradient explosion and gradient disappearance, as well as accelerate the convergence rate for negative values. The overall design has successfully lowered the memory occupancy rate by a substantial margin when combined with the final max-pooling layer.

The multi-head attention mechanism from Transformer [28] is utilized here. This allows us to concentrate on the things that are more relevant. The concepts of Query, Key, and Value need to be introduced in order to determine the correlation. The Key is used to compare with the Query that has to be queried, and after a match is found, the final result is calculated by multiplying it by the Value. Self-attention generates the values of $Q$, $K$, and $V$, i.e., the corresponding linear mapping is done on the most input $z_i$:

$$\begin{cases} Q = z_i * W_Q \\ K = z_i * W_K \\ V = z_i * W_V \end{cases}, \quad (6)$$

here note that the dimensions of $Q$, $K$, and $V$ are the same as the dimensions of $z_i$.

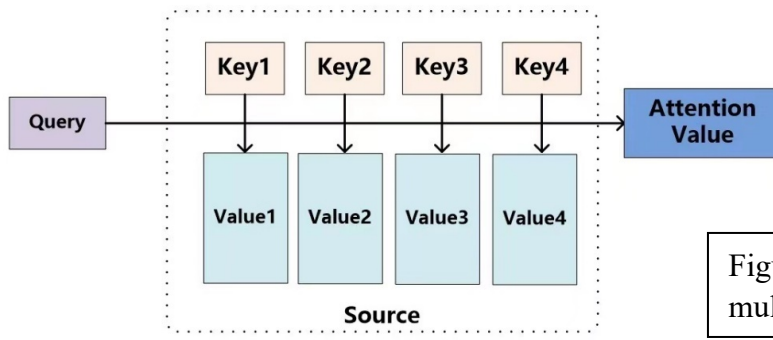

Figure 3: Structure diagram of multi-attentional mechanism.

Multi-head denotes that the $i\text{-}th$ dimension of the matrix is divided into $n$ parts, i.e., $n$ groups of $Q$, $K$, and $V$ are generated by $n$ transformations of $Q$, $K$, and $V$, respectively. It's a module that extracts images' external dependencies. After passing attention, the equivalent output $Head_m$ is generated for each group of $Q_m$, $K_m$, and $V_m$ in this $n\text{-}head$. The final result can be performed by the linear output on all $Head_0 - Head_m$ spliced.

In detail, it may be assumed that $Q$, $K$, and $V$ are one of the heads when computing the correlation. The wider the overlap between the two vectors, the more similar they are, and then $AK^T$ to weight $V$ may be used to produce the attention matrix. The following formula, in the form of a vector dot product, can be used to explain the attention process [28]. Three variables are required for this structure: a Query Matrix $\boldsymbol{Q} \in R^{n \times d}$, a Key Matrix $\boldsymbol{K} \in R^{n \times d}$, and a Value Matrix $\boldsymbol{V} \in R^{n \times d}$:

$$Attention(\boldsymbol{Q}, \boldsymbol{K}, \boldsymbol{V}) = softmax\left(\frac{\boldsymbol{QK}^T}{\sqrt{d}}\right)\boldsymbol{V} = \boldsymbol{AV}, \qquad (7)$$

The correlation between each pair of items is obtained using the dot product operation, and $\sqrt{d}$ is used to convert the attention matrix into a typical normal distribution. After that, softmax is normalized, and the attention score is calculated using the scaling procedure, resulting in a sum of attention weights of 1. As a result, each patch will have all of the information from all other patches, and the dimension of $Attention(\boldsymbol{Q}, \boldsymbol{K}, \boldsymbol{V})$ will match the dimension of $V$. The input sequence could be understood from multiple viewpoints through linear changes after the residual connection and Layer Normalization operations.

Despite the fact that the self-attention mechanism has produced excellent achievements in the realm of medical image semantic segmentation, the large memory capacity required by its quadratic product has become the primary drawback of prediction. Therefore, here the framework of Position Attention [92] will be applied, which will help to considerably lower the time dimension and better cope with longer sequence difficulties when the memory is restricted. To minimize duplicate value combinations in the feature map formed by this mechanism, it is required to extract and optimize dominant features with dominant features, and generate a feature map with a concentrated self-attention mechanism in the following layer.

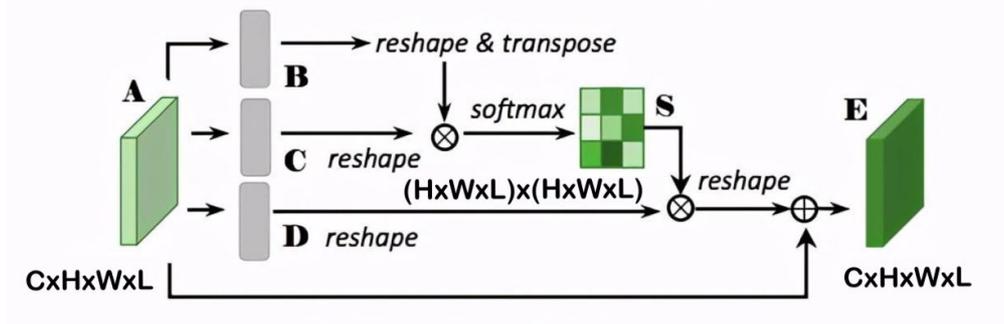

Figure 4: Structure diagram of Position Attention Module (PAM)

First, two tensors of the same shape are created when BatchNorm and PReLU activate the original information flow. The original information tensor is designated by the letters $F \in R^{C \times H \times W \times L}$, whereas the new tensors are designated by the letters $A$ and $B$. Subsequently, $A$ and $B$ are transformed from rank 4 tensors to rank 3 tensors. The purpose is to flatten the original characteristics of $H \times W \times L$ and then transpose one of the periods, resulting in two matrices, $R^{C \times (H \times W \times L)}$ and $R^{(H \times W \times L) \times C}$. When the two are multiplied, the resulting matrix is $R^{(H \times W \times L) \times (H \times W \times L)}$. The softmax procedure is then used to determine the relationship strength between the two location points $i\text{-}th$ and $j\text{-}th$ in the initial attention feature map, yielding:

$$s_{ji} = \frac{\exp(A_j \cdot B_i)}{\sum_{j=1}^{N} \exp(A \cdot B_i)}, \tag{8}$$

Here, the more similar the feature representations of two locations, the greater the correlation between them.

The original feature map $F$ is multiplied by the position attention value at this place. Through the convolutional layer procedure, a new feature map $C \in R^{C \times H \times W \times L}$ with the same shape as F is generated. After reshaping $C$, we obtain $C \in R^{C \times (H \times W \times L)}$, and after transposing, we obtain a matrix that belongs to $R^{(H \times W \times L) \times C}$. Finally, the attention feature map $F_{PAM} \in R^{C \times H \times W \times L}$ will be produced using a weight $\lambda$ that is gradually learned from 0 and element-wise accumulation and summation on the feature $F$:

$$F_{PAM,i} = \lambda \sum_{j=i}^{N} (s_{ji} C_j) + F_i. \tag{9}$$

The weighted total of the features at all places, as well as the original features, makes up the new feature map. At the same time, the weight, which is learned from 0 in a crude manner, may be used to regulate some circumstances when the attention value is too low. As a result, this module is global in scope and capable of combining global context information with learned characteristics.

## 3.3 Network Decoder

Medical image segmentation needs to be implemented in space $(H \times W \times D)$, and a 3D decoder for up-sampling and pixel-level segmentation needs to be introduced now. It is based on the U-Net network architecture [89], which has shown magnificent results in the semantic segmentation of various datasets. The decoder suggested in this article is also inspired by this structure.

Here the multi-resolution feature of the encoder from the Section 3.2 will be merged with the decoder. Different from the original architecture, multi-head cross-attention modules and multi-scale fusion will be carried out after each up-sampling. It is reshaped into the $\frac{H}{P} \times \frac{W}{P} \times \frac{D}{P} \times K$ tensor after extracting the sequence representation $z_i$ with the size of $\frac{H \times W \times D}{P^3} \times K$. Here $K$ represents the embedding size of the encoder. The embedded space from Section 3.2 is represented now, and the $K$ value represents the reshaped feature size. The projection reshaping tensor will be entered into the embedding space for each distinct resolution, and then the upsampling process will be conducted. The HCNN's multi-core convolution operation [88] will be adopted here to try as a feature mapping method to enhance the ability to extract longer sequence features.

### 3.3.1 Attention Gate Blocks

Due to the localization of the receptive field, some faults in the context feature extraction are highlighted during the upsampling phase. Furthermore, Oktay et al. [93] showed that false-positive predictions for tiny objects with substantial form changes reduce accuracy. Then, to properly capture the information, consider adding the attention gate architecture to each upsampling step. The attention gate's construction is depicted in the diagram below:

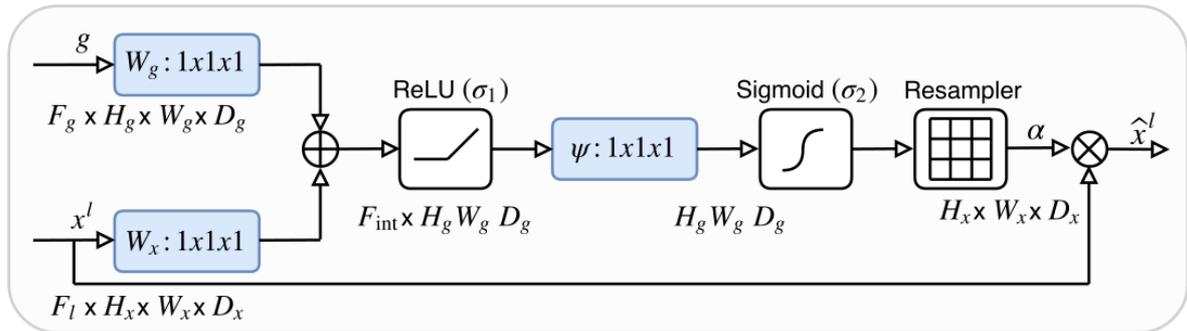

Figure 5: Structure diagram of Attention Gate

The scaling of the coefficients $\alpha_i \in [0, 1]$ in this architectural diagram may be used to reduce significant aspects in the image, enabling only specific tasks to be activated. The output of the Attention Gate is generated by multiplying the input feature map by the relative coefficient for each pixel vector $x_i^l \in R^{F_l}$:

$$\hat{x}_{i,c}^l = x_{i,c}^l \cdot \alpha_i^l, \tag{10}$$

here $F_l$ represents the amount of comparable feature maps at $l\text{-}th$ level. The multi-dimensional attention coefficients indicated in [94] can aid with information embedding based on the present multi-semantic situation. At this time, each attention gate will concentrate on a little portion of the target structure. Each pixel $i$ will have a more accurate gating signal $g_i \in R^{F_g}$ applied to it, which will activate the context information and lower the overall influence of low feature layers, improving classification [95]. The gating coefficients are precisely regulated by additive attention [96, 97]:

$$q_{att}^l = \varphi^T\left(\sigma_1\left(W_x^T x_i^l + W_g^T g_i + b_g\right)\right) + b_\varphi, \tag{11}$$

$$\alpha_i^l = \sigma_2\left(q_{att}^l(x_i^l, g_i; \Theta_{att})\right), \tag{12}$$

where $\sigma_2(x_{i,c}) = 1/(1 + exp(x_{i,c}))$ corresponds to the sigmoid activation function, $\Theta_{att}$ are contained by: the linear transformations $W_x \in R^{F_l \times F_{int}}$, $W_g \in R^{F_g \times F_{int}}$, $\varphi \in R^{F_{int} \times 1}$ and bias terms $b_\varphi \in R$, $b_g \in F_{int}$.

In the sense of vector concatenation-based attention [98], it will execute a linear operation via convolution of channels $1 \times 1 \times 1$ and linearly map the concatenated features $x^l$ and $g$ into the space $R^{F_{int}}$ for an input tensor. In addition, in [93], a grid-attention approach is developed that allows the attention gate to conduct typical back-propagation updates for training: using the gating signal $g_i$ of each skip connection, aggregate information from several imaging scales, where $g_i$ is determined by Network signals for image spatial information rather than a global single vector for all image pixels. This technique considerably improves the query signal's network resolution.

This architecture may emphasize the characteristics of skip connection transfer in the overall MU-Net framework. Specifically, the approximately extracted data may be input through gating to filter out noise, and the combined information can be fed into the concatenation operation via sigmoid. During backpropagation, the activations of neurons in the forward and reverse passes are filtered, and gradients from the background areas are weighted down. This will facilitate the model to update the parameters according to the given task. The following is the update formula for convolution parameters from layer $l-1$ to layer $l$:

$$\frac{\partial(\hat{x}_i^l)}{\partial(\Phi^{l-1})} = \frac{\partial\left(\alpha_i^l f(x_i^{l-1}; \Phi^{l-1})\right)}{\partial(\Phi^{l-1})} = \alpha_i^l \frac{\partial\left(f(x_i^{l-1}; \Phi^{l-1})\right)}{\partial(\Phi^{l-1})} + \frac{\partial(\alpha_i^l)}{\partial(\Phi^{l-1})} x_i^l. \tag{13}$$

The scaling term $\alpha_i^l$ corresponds to a vector at each grid size for multidimensional attention gates. It will be utilized to extract and fuse feature information at different levels in each sub-attention gate, and it will be skip-connected for output.

## 3.3.2 HCNN Structure and Multi-scale Block

The encoder's final layer utilizes the deconvolution layer to alter the feature mapping. In addition, a feature mapping module needs to be created to adapt to the volumetric HCNN [88] decoder's input dimension and transfer the projection reshaping tensor in the embedding space to the input space. The functioning of these skip connections allows the encoder's features to be linked to the decoder's features, resulting in more complex segmentation masks and more extensive spatial information.

Specifically, this series of sequence data must be translated back to a conventional four-dimensional feature mapping, with the resized feature map being linked to the map generated by the preceding encoder. The specific concept is to process the output sequence of the encoder $z_I \in R^{d \times N}$ through the HCNN operation and then reshape it into an array of size $d \times (H \times W \times L)/P^3$.

The convolution block will be utilized to lower the dimensionality of the channel, further lowering the time complexity. Finally, the feature map $Z \in R^{(H \times W \times L) \times C/P^3}$ can be obtained, which dimension is the same as the characteristic map $X$'s dimension in the encoding process. The convolution block will be utilized to lower the dimensionality of the channel to further minimize the temporal complexity.

The HCNN module is built within the Multi-scale Block's framework. Information at different scales could be successfully recorded and reflected in the final output by fusing and extracting the characteristics of each neighboring two layers. The multi-scale module generated after matching will enter the upsampling process to further synthesize and trim the recovered information, using the convolution of different size receptive fields to extract more diversified contextual information. The output of each multiscale module can be determined in the following ways:

$$x_1 = w_{32}(w_{31}input + b_{31}) + b_{32}, \qquad (14)$$
$$x_2 = w_{42}(w_{41}x + b_{41}) + b_{22}, \qquad (15)$$
$$X = Cat[x_1, x_c], \qquad (16)$$
$$Output = w_f X + b_f, \qquad (17)$$

where $input$ denotes the information entered in the previous layer, $x_1$ and $x_2$ represent the features obtained by convolution operations on separate receptive fields, and $Output$ is the outcome of this layer.

## 3.3.3 Feature Up-sampling

Following the feature mapping, $Z$ requires upsampling and convolution procedures. This paper employs the cascaded upsampler approach described in [73] for the former. Specifically, this sampler is made up of multiple up-sampling phases that decode hidden features before returning to the $R^{H \times W \times D}$ space for segmentation masking.

The reshaped full-resolution segmentation result is obtained by first converting the hidden feature sequence $z_I \in R^{d \times \frac{H \times W \times D}{P^3}}$ into the shape of $\frac{H}{P} \times \frac{W}{P} \times \frac{D}{P} \times d$, and then mapping the output from $\frac{H}{P} \times \frac{W}{P} \times \frac{D}{P}$ to $H \times W \times D$ through the cascaded upsampler.

This procedure is continued at consecutive levels until the input resolution reaches the starting value. Each module includes a $3 \times$ upsampling operator, a $2 \times 2 \times 2$ convolutional layer, and a $PReLU$ layer. The $PReLU$ activation function is utilized to feed the result into the $1 \times 1 \times 1$ convolution layer to create pixel-level semantic segmentation.

### 3.3.4 Multi-scale Output

To better encapsulate global and local context information, multi-scale characteristics are considered to be combined. Because the resolutions of the output scale features acquired by each upgraded connection layer differ, the high-efficiency learning characteristics of multi-scale features can forecast and fuse the bilinear difference to get the same resolution value. Finally, these feature maps of distinct sizes need to be integrated to generate a tensor that integrates the deep and shallow detailed information with the semantics of global information. It can be expressed as a multi-scale feature prediction fusion function in the form of convolution:

$$F = fusion([X_0, X_1, X_2, X_3]), \tag{18}$$

## 3.4 Loss Function

This study innovatively combines the Tversky loss [90] and the class-balanced cross-entropy loss [91] as a new loss function for the first time. Tversky Loss can employ classic dice to further balance false positive and false negative parameters within the network to better detect tiny lesions in imbalanced data. For cross-entropy, as a fundamental loss function, can be a decent indicator of how effectively a classifier can categorize lesions. Therefore, when paired with the optimizer, the combination of these two loss functions can converge to the global minimum more thoroughly and precisely than a single loss function. The pixel-level equations can be used to derive it:

$$Loss(G, P) = 1 - \frac{\sum_{i=1}^{I} P_{i,j=0} \, G_{i,j=0}}{\sum_{i=1}^{I} P_{i,j=0} \, G_{i,j=0} + \alpha \sum_{i=1}^{I} P_{i,j=0} \, G_{i,j=1} + \beta \sum_{i=1}^{I} P_{i,j=1} \, G_{i,j=0}}$$

$$-\lambda \sum_{m=0}^{3} \left\{ \sum_{j=1}^{J} \sum_{i=1}^{I} \left[ G_{i,j} log(P_{i,j}) + (1 - G_{i,j}) log \left( 1 - log(P_{i,j}) \right) \right] \right\}, \quad (19)$$

where $I$ denotes the voxels' number; $J$ denotes the classes' number; $P_{i,j}$ means the probability of the $i\text{-}th$ voxel and the $j\text{-}th$ class; $G_{i,j}$ means the one-hot encoded ground truth of the $i\text{-}th$ voxel and the $j$-th class.

Note that the loss function component of the cross-entropy based on the multiscale output is represented by the third term. This is a specific instance of multi-category cross-entropy and can be regarded as the cross-entropy of two categories.

## 3.5 Theoretical Limitations

Through the theoretical analysis in Section 3.1-3.4, it can be found that a 3D image will be reconstructed into a sequence and then fed into the Position Attention Module for enhanced feature extraction. Between the first and third stages of the encoder, the attention mechanism is not used until this level is achieved. On the one hand, the grasping of information in the shallow network is not strengthened under the attentional mechanism. On the other hand, the image is missing information after three times of downsampling. This was tried in the experiment, but unfortunately, the results were disappointing, so we only employed one PAM layer in the MPU-NET architecture. Although this design outperformed U-Net in the experiment, more efforts need to be done in the future to add a position attention module in each layer of the encoder to properly extract and de-sample the feature information of each class.

Moreover, at the multi-scale output level, the fusion operation might have a variety of options. The simplest weighted average, which is the most conservative approach, is adopted in further tests using this design. We also attempted using the Pyramid Pooling Module [107] to produce the final segmentation output. However, the results achieved with multi-scale attention blocks were not adequate. Therefore, we reduce the intricacy in this paper. This conservative decision may not increase the framework's overall performance, but it does take into account feature fusion at various sizes. We may try to build this PPM layer under the assumption of other attention methods in the future, and broaden the receptive area to better utilize the global context information.

# 4 Experimental Setup

## 4.1 Database and Data Preprocess

We employ a 3D medical image segmentation dataset based on CT imaging modalities to test the suggested model - LiTS (Liver Tumor Segmentation Challenge). This database includes 130 CT scans for training and 70 CT scans for testing. The boundaries of the liver were delineated and annotated by researchers Bilic et. al [100], funded by the International Conference on Medical Image Computing and Computer Assisted Intervention (MICCAI, 2017). This dataset (LITS) is publicly available as part of ISBI 2017. Based on this collection of data, this work attempts to perform automatic segmentation of liver tumors.

Table1: Division of patient samples.

|  | Training | Validation | Testing | Total |
|---|---|---|---|---|
| CT Scans | 88 | 22 | 20 | 130 |

Due to authorization difficulties, we did not get the complete 200 groups of patients' data. Therefore, we divided the 130 CT scans into a training group, a validation group, and a test group at random in a ratio of 88:22:20. The pixels of the cross-section of the image is 512x512, and the number of slices contained in each set of images is variable, ranging from less than 50 to around 1000. The ground truth segmentation in the LiTS dataset gives three separate labels for the liver, tumor lesion region, and backdrop. By utilizing image preprocessing, we were able to remove those scans below 48 slices and classify the liver as positive and others as negative.

For the experiments, each volume was independently preprocessed. A simple data augmentation strategy is performed during preprocessing where we randomly rotate, flip, crop, and mirror images without increasing the image size. Before being trained by the model described in this study, the input raw images are downsampled and scaled to 256x256 pixels for liver tumor segmentation. The model's input image is also 256x256 pixels in size. Our technique will increase the efficiency of judging liver tumors in abdominal CT scans by describing the liver tumor segmentation issue as a binary classification segmentation assignment with one channel number.

## 4.2 Baselines and Implementation

Different from the standard U-Net segmentation design, we propose a 3D automatic segmentation algorithm to collect global contextual data. The gradient update is calculated in tiny batches with $batch\ size = 1$ because the gradient is averaged across numerous forward and backward passes, and the computational power & video memory requirements are too large. Our automatic segmentation algorithm uses the Adam optimizer [102] and batch normalization to train the segmentation objective function (the loss function) mentioned in Section 3.4. At the end of the training, the lossy Softmax will be applied to compare the output of the network with the labels of the ground truth. The 3D abdominal CTs of all patients were trained slice by slice, and each slice was then rebuilt into 3D for assessment.

We use the Python3 programming language for experiments and utilize the PyTorch (version 1.8.0) framework proposed by Paszke et al. [101] to develop all models, running on an NVIDIA Quadro RTX 6000 GPU with 22GB of RAM. The initial learning rate $lr$ is 1e-4, the momentum is 0.1, and $lr$ is halved after every 100 iterations. The total number of iterations for each training varied between 50 and 600. After each training session, it will be compared with the previous model, and it will be stored if it performs better. Each loop takes roughly 2 minutes on average. The aforementioned settings were chosen after several testing, and they can help the loss function approach and limit more efficiently.

## 4.3 Evaluation Metrics

Six assessment criteria will be utilized to examine the effectiveness of the medical image semantic segmentation model for liver tumor segmentation. These metrics are commonly employed in image segmentation tasks, and they will aid us in evaluating the performance and quality of various models during training, validation, and testing.

Table2: The confusion matrix

| Confusion matrix | | Predict | |
|---|---|---|---|
| | | 1 | 0 |
| Real | 1 | True Positive | False Negative |
| | 0 | False Positive | True Negative |

We first create the above matrix. Here, we can let the abbreviation TP represent True Positive: the size of the true tumor region in pixels; TN denotes True Negative: the size of the true non-tumor region in pixels; FP denotes False Positive: the size of the false tumor region in pixels; FN denotes False Negative: the size of the false non-tumor region in pixels.

Moreover, $G_i$ and $P_i$ can be used to signify the ground truth and prediction results of voxel $i$ respectively based on the given semantic class. Then these metrics can be defined by:
1) Dice coefficient: It is an ensemble similarity measure function that measures the overlapping part of two samples. The index ranges from 0 to 1, with 1 indicating total overlap. It's also known as the F1 score:

$$Dice(G, P) = \frac{2\,|Ground\ Truth \cap Segmentation|}{|Ground\ Truth| + |Segmentation|}$$

$$= \frac{2\sum_{i=1}^{I} G_i P_i}{\sum_{i=1}^{I} G_i + \sum_{i=1}^{I} P_i}$$

$$= \frac{2\,TP}{2TP + FP + FN} \tag{20}$$

2) Accuracy coefficient: This is a popular assessment metric that divides the total number of samples by the number of correct prediction outcomes:

$$Accuracy = \frac{TP + TN}{TP + TN + FP + FN} \tag{21}$$

3) Precision coefficient: It is an accuracy metric that represents the percentage of accurately predicted positive classes to the total number of positive classes anticipated:

$$Precision = \frac{TP}{TP + FP} \tag{22}$$

4) Specificity coefficient: It gauges the classifier's ability to detect negative instances by comparing the size of accurately predicted negatives to the total amount of real negatives:

$$Specificity = \frac{TN}{TN + FP} \tag{23}$$

5) IOU coefficient: It is calculated by dividing the overlapping area of the labels A and the predicted outcomes B by the combined area of A and B. It is used to assess the similarities and differences between small sample sets and is concerned with whether the qualities shared by individuals are consistent:

$$IOU = \frac{TP}{TP + FP + FN} \tag{24}$$

6) Matthews correlation coefficient (MCC): This is one of the evaluation indicators for evaluating the binary classification model's outcomes, and it may help address the indicator measurement problem of unbalanced category data:

$$MCC = \frac{TP \times TN - FP \times FN}{\sqrt{(TP + FP)(TP + FN)(TN + FP)(TN + FN)}} \tag{25}$$

The first five assessment criteria have a value range of 0 to 1. The larger the coefficient value, the wider the overlapping region between the segmentation image and the ground truth, and the better the model effect. For MCC, it gives a number between -1 and +1. The coefficient +1 denotes flawless prediction, 0 denotes no better than chance, and -1 denotes full contradiction between forecast and observation.

# 5 Results and Analytical Studies

## 5.1 Segmentation Experiments & Basic Analysis

We perform several tests and analyze the experiment results from various perspectives in order to further assess the automated segmentation framework described in this work. As described in Section 4.1, 88 patient samples will be randomly selected for training during training, and then 22 samples will be validated. Six evaluation metrics described in 4.3 can be reported. This section will present and evaluate the longest experimental record to better demonstrate the model's effectiveness. Experimental records kept: the initial learning rate $lr$ is 1e-4, which is lowered by a factor of $1/2$ every 100 iterations. This experiment had a total of 600 iterations at the end. The experimental training time is 20 hours. The comparison of the six assessment criteria during the training and validation processes is shown in the following six pictures:

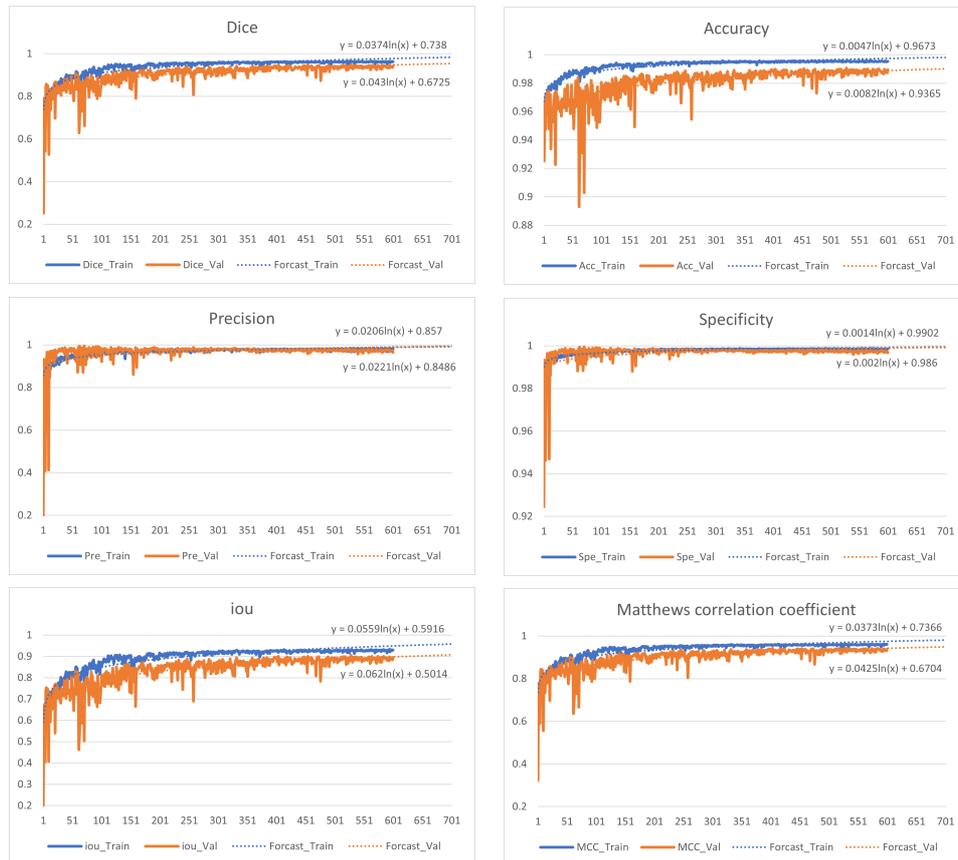

Figure 6: Visual comparison of metrics for training and validation of MPU-Net.

We can observe from these 6 sets of graphs that, while each evaluation indicator tends to converge after a given number of repetitions of training, there is still space for improvement in each indicator. The histogram below helps us see the indicator's existence interval more clearly.

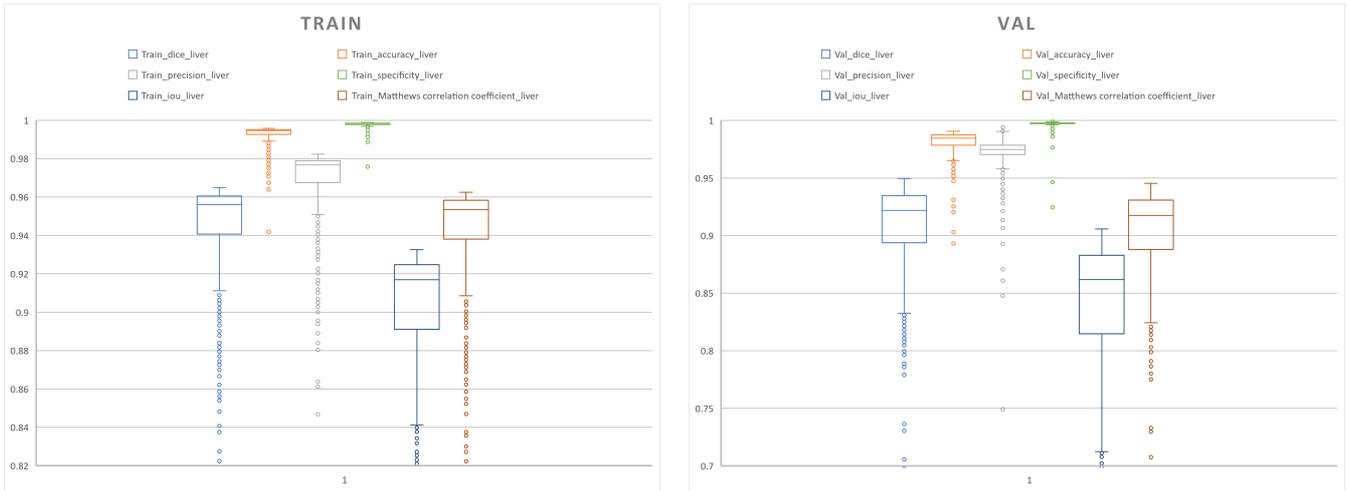

Figure 7: Boxplots of metrics for training and validation of MPU-Net.

It's easy to observe from these two graphs that the median of each indicator ranges from 0.85 to 1. The accuracy coefficient, precision coefficient, and specificity coefficient all surpass 96% in the training set. These three indicators are also essentially more than 95% for the validation set. This suggests that the present model can accurately anticipate samples and has a high rate of positive sample identification. Furthermore, we can see that after 200 iterations, each observation index begins to converge through the above 6 graphs. One of the causes is that when the iteration is 200, $lr$ drops by a factor of $0.5^2$: $lr = 2.5e - 05$. However, there is certain evidence that reveals a very unequal fallback between 200 and 400 iterations. To be more exact about each indicator's ultimate limit, we compute the average of each indicator from 400 iterations to the last 600 iterations, yielding the table below. For a more obvious comparison, a clustered bar chart of training and validation may be created:

Table 3: The convergence data for training and validation of MPU-Net.

|  | Dice | Accuracy | Precision | Specificity | IOU | MCC |
|---|---|---|---|---|---|---|
| Train | 0.960792 | 0.995213 | 0.978838 | 0.998553 | 0.9255525 | 0.9584755 |
| Val | 0.934909 | 0.9873545 | 0.973139 | 0.997362 | 0.8830005 | 0.9308275 |

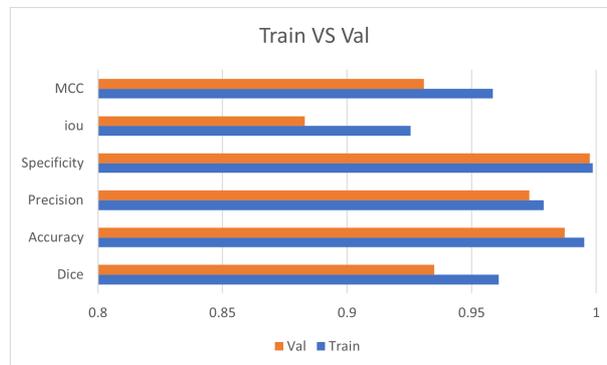

Figure 8: Comparison histogram of metrics for training and validation of MPU-Net.

We next conduct experiments with U-Net and record the results. Each of the following six indicators has been in a stable state, indicating that this model has limited opportunity for improvement.

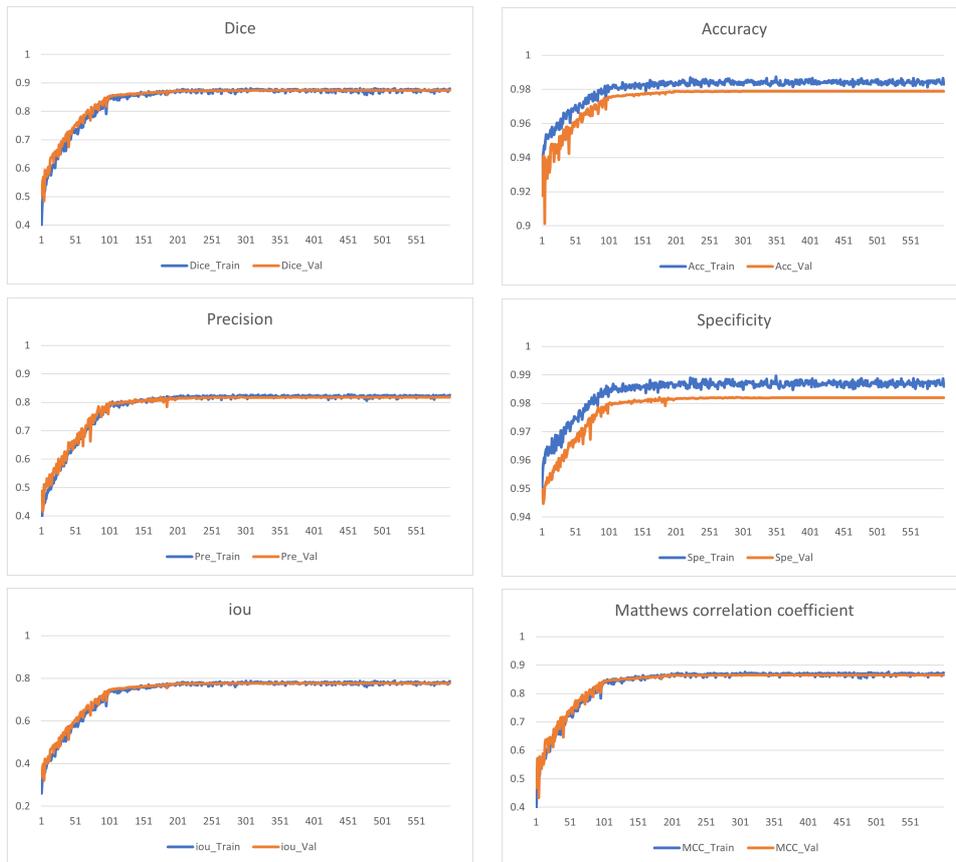

Figure 9: Visual comparison of metrics for training and validation of U-Net.

Similarly, the following box plots will allow us to observe the interval range of each indication more clearly.

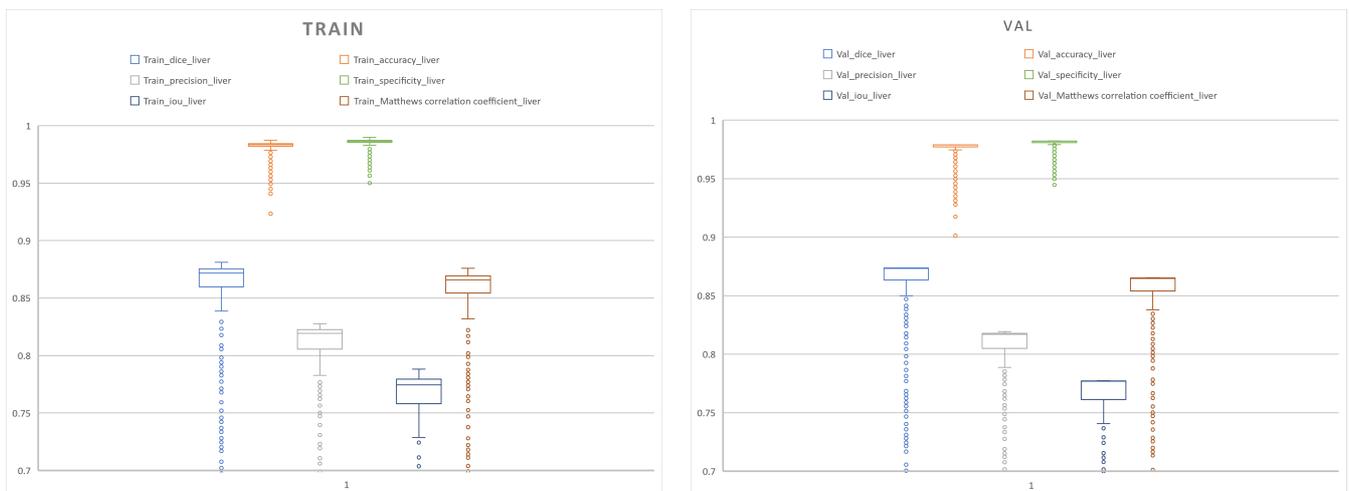

Figure 10: Boxplots of six different metrics for training and validation of U-Net.

The box plot clearly shows that the traditional U-Net model has shown excellent performance under some evaluation criteria. For example, the accuracy and specificity indicators are comparable to the MPU-Net. The remaining four characteristics, on the other hand, have a value range of roughly 70% to 85%. It is clear that the model proposed in this paper occupies a strong position in several areas. However, unlike MPU-Net, practically every indication in the U-Net model does not fall back after 200 iterations. The average of 400-600 iterations of U-Net is also determined for a fair comparison, and the following table is generated, along with a clustered bar chart:

Table 4: The convergence data of metrics for training and validation of U-Net.

|  | Dice | Accuracy | Precision | Specificity | IOU | MCC |
| --- | --- | --- | --- | --- | --- | --- |
| Train | 0.873704 | 0.984101 | 0.820985 | 0.986936 | 0.777716 | 0.867728 |
| Val | 0.8736 | 0.9789 | 0.817744 | 0.982 | 0.777211 | 0.865111 |

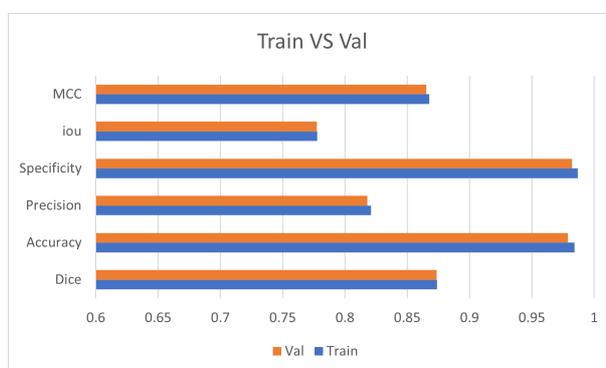

Figure 11: Comparison histogram of metrics for training and validation of U-Net.

The next section provides a more detailed evaluation and comparison of the properties of each model.

## 5.2 Qualitative evaluation and comparison

From the 6 sets of indicator data charts about MPU-Net shown in Section 5.1, it can be found that the 6 indicators are all extremely close to 1. Accuracy, Precision, and Specificity perform the best, with all three above 97 percent, and Accuracy and Specificity even approaching 100 percent. This indicates that this model's segmentation ability is excellent. However, the histogram shows that the performance of the IOU indicator is marginally lower in comparison. This might be due to a lack of feature extraction in the supplied image. Recall that the IOU metric represents the similarity of two sample sets. During training, the average of this metric is 3%-7% lower than the other evaluation criteria; in the validation set, it is also 5%-11% lower than the other criteria. Despite being marginally degraded, this metric can achieve a maximum of more than 88%.

It also performs admirably in Accuracy and Specificity for the standard model U-Net. In the U-Net experiment, these evaluation values reveal above-average model performance when compared to the assessment in [60]. This should be the benefit offered by the loss function because the U-Net experiment uses a new loss function presented in this paper: a mix of Tversky loss and cross-entropy loss. We may also compare the two models by comparing the differences in each parameter. When we subtract the value in MPU-Net from the value in U-Net, we get:

Table 5: Comparison of the 6 metrics of MPU-Net and U-Net in the training set and the validation set.

|  |  | Comparison | | | | | |
| --- | --- | --- | --- | --- | --- | --- | --- |
|  |  | Dice | Accuracy | Precision | Specificity | IOU | MCC |
| MPU-Net | Train | 96.08% | 99.52% | 97.88% | 99.86% | 92.56% | 95.85% |
| | Val | 93.49% | 98.74% | 97.31% | 99.74% | 88.30% | 93.08% |
| U-Net | Train | 87.37% | 98.41% | 82.10% | 98.69% | 77.77% | 86.77% |
| | Val | 87.36% | 97.89% | 81.77% | 98.2% | 77.72% | 86.51% |

The results demonstrate that the architecture described in this report outperforms the benchmark model U-Net in every metric, and is superior to U-Net in the training set and the validation set of each indicator by 8.71%, 1.11%, 15.78%, 1.17%, 14.79%, 9.08%; 6.13%, 0.85%, 15.54%, 1.54%, 10.58%, 6.57%, respectively. Among them, the performance of MPU-Net is not much better than the benchmark model U-Net in terms of accuracy and specificity. One argument is that this model has too many parameters, resulting in too much duplication in the forward and backward transfer of variables. The resulting mistake will likewise rise. But overall, the capabilities revealed by MPU-Net show that our proposed architecture can conduct volumetric segmentation of patient 3D pictures with greater accuracy.

In addition, we can analyze and compare the two models from the perspective of convergence and computation time. The average time it takes for the two models to finish a single training session is listed in the table below. The calculation time includes everything from loading the preprocessed 3D slice pictures into the model to reading and training each image and slice, as well as keeping the current parameters. Both architectures are tested on the devices mentioned in Section 4.2.

Table 6: Comparison between MPU-Net and U-Net.

| Method | Train_time | Val_time | Output | Total time | Parameters | File Size |
| --- | --- | --- | --- | --- | --- | --- |
| MPU-Net | 107s | 12s | 1s | 120s | 22557009 | 250.34MB |
| U-Net | 17s | 3s | 1s | 21s | 2326638 | 26.65MB |

The parameters of the method described in this article are around 10 times those of the basic model U-Net architecture, implying that the model will demand more processing power during training and verification. Note that during the training phase, the benchmark framework U-NET may set batch size = 10, but as MPU-Net can only use batch size = 1, the standard model tests set batches of the same size to compare the performance of these two models generated in a single session. Numerous gradients propagated forward and backward are included in the modules built using the attention mechanism, resulting in a considerable increase in the total computational complexity of the model, with a computational time improvement of nearly 6 times. Another aspect of having more parameters is that MPU-Net outperforms U-Net.

Moreover, we can compare the value $1 - dice\ scores$ of the two segmentation algorithms together to generate the following figure in terms of convergence. We discover that the network's end conclusion is quite comparable. In some ways, the convergence behavior of the two models is similar.

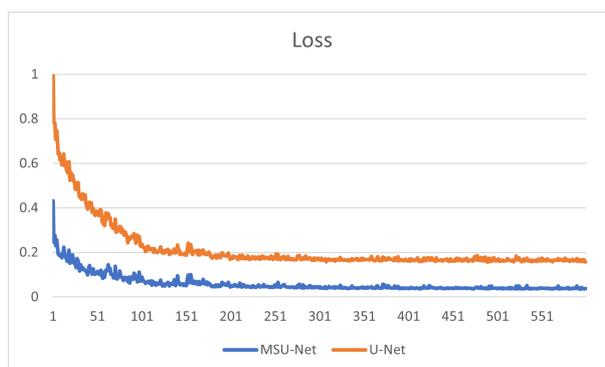

Figure 12: Visual comparison of convergence process between MPU-Net and U-Net.

Because the learning rate was considerably lowered, both models began to show a progressive convergence tendency after 100 iterations, as seen in the above figure. The fundamental rate of change of their loss values has stabilized after 200 iterations. The value of loss reduces more slowly as the number of repetitions rises. The best results produced by MPU-Net and U-Net are both above the standard performance in medical image segmentation, with the greatest results coming in at 528 and 390 times, respectively. We may extract these two sets of data independently for comparison, resulting in the table below:

Table 7: Performance comparison of the best-saved models of MPU-Net and U-Net.

| | The best model of each framework | | | | | |
|---|---|---|---|---|---|---|
| | MPU-Net | | U-Net | | MPU minus U | |
| Iterations | 528 | | 390 | | | |
| | Train | Val | Train | Val | Train | Val |
| Dice | 96.45% | 94.94% | 87.5% | 87.37% | 8.95% | 7.57% |
| Accuracy | 99.51% | 99.07% | 98.43% | 97.89% | 1.08% | 1.18% |
| Precision | 98.24% | 96.65% | 82.26% | 81.73% | 15.98% | 14.92% |
| Specificity | 99.85% | 99.68% | 98.71% | 98.2% | 1.14% | 1.485 |
| IOU | 93.18% | 90.58% | 77.94% | 77.74% | 15.24% | 12.84% |
| MCC | 96.2% | 94.53% | 86.9% | 86.53% | 9.30% | 8% |

By comparison, those main metrics for MPU-Net are shown to be superior to those for U-Net. Therefore, the MPU-Net architecture shown in this paper outperforms the competition in both detail and overall performance, approaching 100% in accuracy and specificity assessments. In terms of dice and MCC, U-Net is much inferior to MPU-Net, although doing well on the other two criteria.

The image segmentation effect based on this architecture is demonstrated below, based on the good performance comparison of MPU-Net. We utilize ITK-SNAP software to visualize the nii file format. The best model saved throughout the training was chosen as the parameters that were kept at the 528th iteration. The segmentation results of the 3D volume picture of the first patient identified as NO.27 were randomly picked from the test set for visual display.

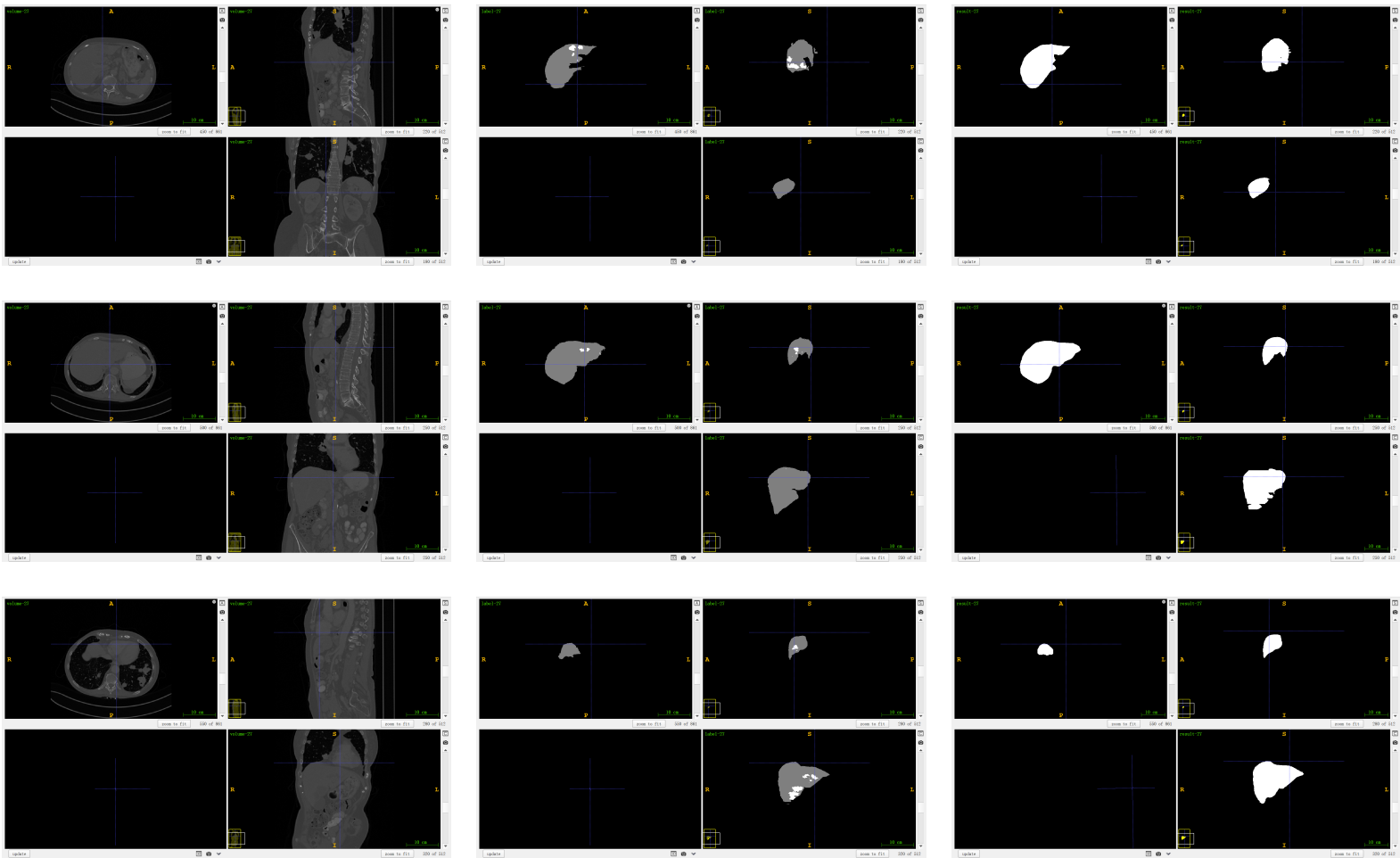

Figure 13: Graphical representation of the MPU-Net model's segmentation results.

This is a set of 512x512x861 pixel nii patient CT images. We choose three slices at random for display, and their coordinates are: (220, 180, 450), (250, 250, 500), (280, 320, 550). Note that there are no abdominal tumors in many coordinates, and the coordinates we chose here are all within the range where the tumor is located. In the above nine pictures, the first column is the original CT scans, the second column is the result of expert manual segmentation, and the third column shows the segmentation outcome of the parameters that match the best parameters of MPU-Net. It should be noted that the coordinates for the segmented nii file need to be converted to RPI through the Reorient Image operation in ITK-SNAP. Because we employed typical data augmentation techniques to rotate the axial, sagittal, and coronal planes of the original 3D picture at random during training. The goal of this operation is to prevent the model from discriminating during training by using a fixed location. This procedure improves the model's generalization capabilities. It may be adjusted appropriately by "zoom to fit" to acquire the results of the size standard.

The impact reveals that the MPU-Net segmented tumor portion in the picture is nearly identical to the tumor label. Note that throughout the training phase, the normal distribution's intensity values are linearly scaled n via an affine transformation. The final results are saved in the folder: \MPU-Net_Results\Segmentation_Results. Some of the photos in the other 19 nii files are compressed, giving the impression that the organs are flattened. This is normal because the LiTS2017 dataset we downloaded has this feature. The tumor labels that the doctor had tagged are saved in the folder: \MPU-Net_Results\Compressed_Labels. The nii.gz file, which has a size of 2.51MB (The original file is 864MB in size), was compressed using the MATLAB's function gzip('nifti-file-name.nii'). Because each 3D patient picture is compressed once, it is typical to see that the tumor borders in the label do not appear smooth.

The following table is then used to intuitively observe the segmentation effect of 3D patient CTs in 20 test sets.

Table 8: Performance data presented in test sets for the best model saved by MPU-Net.

| img | Dice | Accuracy | Precision | Specificity | IOU | MCC |
|---|---|---|---|---|---|---|
| 27.nii | 0.9356 | 0.9982 | 0.899 | 0.9985 | 0.879 | 0.9354 |
| 28.nii | 0.9411 | 0.9896 | 0.9802 | 0.9981 | 0.8887 | 0.9363 |
| 29.nii | 0.9569 | 0.9968 | 0.9657 | 0.9987 | 0.9173 | 0.9553 |
| 30.nii | 0.952 | 0.9958 | 0.9277 | 0.9966 | 0.9084 | 0.9502 |
| 31.nii | 0.9232 | 0.9922 | 0.9856 | 0.9993 | 0.8573 | 0.9211 |
| 32.nii | 0.9602 | 0.9954 | 0.9581 | 0.9974 | 0.9234 | 0.9577 |
| 33.nii | 0.9537 | 0.9953 | 0.9333 | 0.9964 | 0.9114 | 0.9514 |
| 34.nii | 0.9241 | 0.9943 | 0.9414 | 0.9977 | 0.859 | 0.9213 |
| 35.nii | 0.7368 | 0.9439 | 0.7941 | 0.977 | 0.5833 | 0.7079 |
| 36.nii | 0.9583 | 0.9963 | 0.9767 | 0.9989 | 0.92 | 0.9566 |
| 37.nii | 0.9618 | 0.9955 | 0.9583 | 0.9974 | 0.9263 | 0.9594 |

| 38.nii | 0.9343 | 0.9936 | 0.9368 | 0.9968 | 0.8766 | 0.9309 |
| 39.nii | 0.8969 | 0.9902 | 0.8674 | 0.9932 | 0.8131 | 0.8924 |
| 40.nii | 0.9667 | 0.9957 | 0.9635 | 0.9975 | 0.9355 | 0.9644 |
| 41.nii | 0.937  | 0.993  | 0.9451 | 0.9968 | 0.8815 | 0.9334 |
| 42.nii | 0.9456 | 0.9955 | 0.9228 | 0.9966 | 0.8969 | 0.9436 |
| 43.nii | 0.8187 | 0.9776 | 0.7552 | 0.9827 | 0.693  | 0.8101 |
| 44.nii | 0.9503 | 0.9933 | 0.9423 | 0.9958 | 0.9053 | 0.9468 |
| 45.nii | 0.9108 | 0.9891 | 0.8853 | 0.9923 | 0.8361 | 0.9054 |
| 46.nii | 0.8702 | 0.9955 | 0.8435 | 0.9971 | 0.7702 | 0.8683 |

In these 20 sets of 3D patient data, we observe that the trained MPU-Net model performs admirably. Four groups failed to attain more than 90% accuracy on the most essential dice coefficient. We can calculate the average of each parameter and make a radar chart to visualize the observation:

Table 9: The average of the six metrics in the test set.

|     | Dice    | Accuracy | Precision | Specificity | IOU      | MCC      |
| --- | ------- | -------- | --------- | ----------- | -------- | -------- |
| AVE | 0.92171 | 0.99084  | 0.9191    | 0.99524     | 0.859115 | 0.917395 |

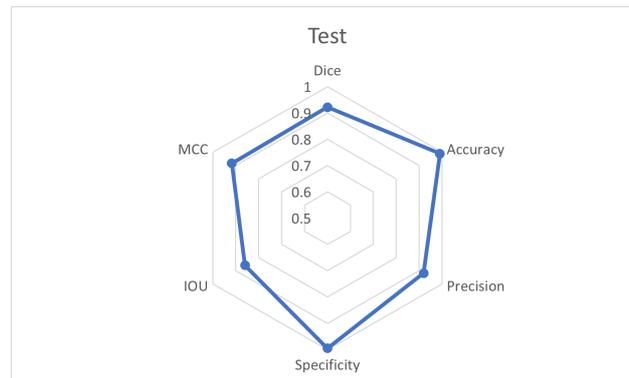

Figure 14: Radar charts for the six metrics in the test set.

The properties of the six indications that emerged in the training are likewise shown in the test set. Although the accuracy and specificity indicator can have a near-100% effect, the IOU is not. The IOU evaluation standard refers to the outcome obtained by dividing the overlap of the two zones of predict and label by their set. Note that the overlap of the predicted and labeled areas divided by their set yields the IOU assessment criterion. The model's generalization and comprehensiveness may be considerably enhanced if this indicator improves. In the next part, the model's potential flaws are investigated and discussed.

# 6 Discussion

## 6.1 Generalizations

On the LiTS dataset, we found that the suggested model outperforms the traditional 3D model U-Net framework in many aspects. In particular, the design of this article has substantially increased the capacity to acquire global information using 3D U-Net. Furthermore, by the end of the training, all indicators are essentially steady, and most of them achieve the optimal value of more than 90%.

For its generalization capabilities in the task of 3D automated tumor segmentation, the model in this research may be evaluated on additional medical datasets. For testing and comparison, "Medical Segmentation Decathlon (MSD)" [31] and "Beyond the Cranial Vault (BTCV)" [32], as well as other image formats such as MRI, can be employed. Furthermore, the TransUNet model [73], which is an effective semantic segmentation approach, might be comparable to this model. Because 3D medical image segmentation technology is still in its infancy, there are not many open sources and effective PyTorch-based frameworks for volumetric medical image segmentation.

The capacity to split data into distinct dimensions is another feature of generalization. The patient's 3D volume scans may be separated into slices individually since the 3D segmentation model is trained on stacking 2D models. What has to be altered in the model code is the convolution dimension operation. The kernel size and padding may also be tweaked as needed depending on the circumstance. At present, there are numerous standard 2D segmentation models, which can be used to compare and evaluate the effectiveness of the MPU-Net architecture with them more widely.

## 6.2 Errors and limitations

First and foremost, by viewing the training process in the experiments in Section 5, we can notice that the model is quite unstable in the early iterations. Except for the first dice coefficient, which is less than 30% owing to parameter setup, it is frequently near 50% in the succeeding iterative operation procedure. To some extent, this reflects the architecture's susceptibility to parameter changes and its low anti-interference factor. Furthermore, close graphical monitoring of the early indicators reveals that even when the model has achieved a decent dice value, there is often a minor negative tendency. Because of the enormous number of training epochs, we cannot see this in the visualization, but it might imply that the model is unstable in its early stages. This might be due to insufficient hyperparameters or a failure of the pre-training model to extract and neutralize information from the attention module while capturing global context data. We may consider this from the standpoint of the optimizer. Optimizer

Adam [17, 103] has been found to have benefits in medical image analysis, while SGD [104] and AdamW [105] may also perform fantastic work for autonomous 3D tumor segmentation tasks when utilizing related algorithms [73,93,99]. Due to time restrictions, we could not test additional optimizers in conjunction with the present model. We will attempt to perform more tests in the future to assess the convergence effect of each optimizer on various datasets.

It is worth mentioning that the PolyLoss function whose coefficients can be modified for different models and workloads was presented in the current study by Google's Leng team [108]. This discovery improves performance and accuracy in a variety of deep learning tasks. Since this was just released publicly on May 1st, we haven't had time to examine the effect of this loss function on volumetric medical image segmentation. We will try this in subsequent research.

Furthermore, the choice of learning rate is also a major issue. The model's learning rate is half every 100 iterations in the present model. However, it was discovered during training that with the present learning rate, the model can reach a stable value after hundreds of rounds. The setting of 100 iterations at this point severely lowers the model's performance, bringing the learning rate at which the model can't advance swiftly to the next stage even closer to the limit. Therefore, it is appropriate to try to include a function to evaluate whether the limit value has been achieved at the current learning rate before moving on to the next training stage during the training and verification process. In addition, visualization reveals that after 100 iterations, the IOU evaluation criterion does not retain a greater rate increase. This may be accomplished by increasing the iteration number of $lr$ in the first phase, but the model will converge to the ideal outcome more slowly. It's also a challenge to think about avoiding making the learning rate iterate too long in the present state and improving the IOU coefficient's convergence rate.

In addition, we noticed that there are thousands of parameters, and the model structure is not sufficiently simplified. This software has about 20 million parameters and is somewhat large to execute, making it impossible to finish an experiment without a server with a high-end NVIDIA graphics card. Even with an attention mechanism module capable of capturing contextual information, this may lead the network to fail to display quick convergence in the early stages. In fact, this experiment attempts to add a Long short-Memory module [69] to improve the capacity to collect long-range information, but due to the fact that the dice value after 50 repetitions still cannot surpass 60%, we have decided to abandon this approach. Clearly, the existing network design is not properly optimized, and module communication is inadequate. Although the acquisition and transmission of contextual information have superior effectiveness when compared to the normal U-Net architecture, it still falls short.

Finally, as indicated in Section 3.5, the ensemble of positional attention blocks at each layer of the encoder has not been realized in the theoretical design or practice of the model. Furthermore, while the present model's ultimate output is based on a weighted average, the way by which the pyramid pooling module balances significant and irrelevant data is worth investigating further. Pytorch-based code hits a bottleneck in achieving both. Perhaps including these two elements can help to enhance the model's overall performance, which will be part of our follow-up study. The next part will outline additional research directions for addressing the model's present issues and achieving breakthroughs.

## 6.3 Future directions

The model's capture of long-range information improves by adding the position attention module. However, the Informer architecture is now more efficient and less mature. When compared to other architectures, the Transformer model takes the lead in the medical image semantic segmentation algorithm. Nevertheless, for the Long Sequence Time-Series Forecasting (LSTF) problems, the Informer structure has emerged as the best option [30]. Informer has become the most powerful sequence prediction magic. It overcomes many possible difficulties in the Transformer architecture. To process extended sequences, it fully utilizes the self-attention mechanism, and the newly developed ProbSparse Self-Attention efficiently minimizes the complexity of quadratic time and memory usage.

For these reasons, the Informer architecture was presented by Zhou et al. [30], which not only solves these concerns but also predicts long-term sequences with only one forward operation, considerably improving computing efficiency. The capacity to efficiently capture the long-distance dependency between output and input is reflected by this flexible expression. This approach can reduce the input of the cascade layer and effectively support longer input sequences.

In truth, the project's initial research aim was to include ProbSparse Self-Attention into the automated segmentation network, but the experiment proved unsuccessful. The appendix contains a study of this attention process. Informer will be a great hit in future medical image segmentation challenges as a better method than Transformer. At this time, few academics have attempted to integrate the Informer architecture into the U-Net network. We will continue to work on improving the learning of this attention mechanism and demonstrating the Informer's efficacy in medical image semantic segmentation. Perhaps the restrictions indicated in section 6.1 can be effectively addressed by using the Informer encoder.

Additionally, following the Multi-head ProbSparse Self-attention module, we may try adding the Bi-LSTM module to strengthen the neural network's capacity to overcome the long-term dependency problem. This will improve the network's ability to capture information and avoid gradient explosion or disappearance.

# 7 Conclusion

This research examines the original U-fully Net's convolutional neural network architecture [16] and offers a multi-scale MPU-Net model based on multi-head cross-attention and position attention mechanisms [92]. MPU-Net can separate liver tumors from patient volume data automatically and reliably. This effective artificial segmentation method can help clinicians make better tumor placement decisions by collaborating with early screening and increasing the success rate of tumor therapy [2, 3]. In the realm of image analysis, this approach is useful for collecting feature information from different layers and for achieving efficient communication with tumor localization and segmentation in clinical practice [24]. The real value of this approach is greater than statistical-based methods [62], and the completely automatic segmentation mode makes the FCN architecture [60, 63], to which MPU-Net belongs, dominate in medical picture semantic segmentation. U-Net has demonstrated higher performance in 2D image segmentation in practice as the current benchmark model in medical imaging [16-20]. However, in the realm of 3D, this technology is still in its infancy. Implementing 3D picture segmentation can give more efficient and accurate clinical support [64], and multiple forms of the attention mechanism [27, 71, 78] can be included in a 3D volume segmentation network due to the flexibility of U-Net. This will allow the performance of each module to be combined, and it will serve as an excellent bridge between 2D and 3D semantic segmentation.

In the testing phase, we illustrate the superior performance of the MPU-Net model in liver tumor segmentation through 6 evaluation metrics on the LiTS dataset. This massive boost in efficiency is attributed to the combination of different attention mechanisms and the balanced consideration of the scale output of each layer, which allows the target can be accurately located while undergoing training. Different semantic information associated between modules is conveniently collected, and the diversity of characteristics is refined. Moreover, a mixed loss function that combines Tversky loss with pixel-level cross-entropy loss is modified in this research to better capture the emphasis from the mixing features and achieve more efficient convergence. Meanwhile, the decoder's performance in feature capture and fusion is improved due to the integration of Hierarchical Convolutional Neural Networks with multi-scale blocks. There are still a lot of literature studies on the application of deep learning to semantic segmentation of 3D medical images, with the goal of integrating multiple attention modules with more accurate and faster convergence. Currently, MPU-Net has not further verified its scalability. Its performance on various organs and tissues may differ, and the instability of the training process in the early stages is another challenge that the model has yet to address. Within this research area, transformer-based topologies have been attempted to maximize usage [73, 81, 84], and future research might focus on incorporating Informer modules in typical 3D semantic segmentation networks.

Furthermore, the structure of MPU-Net is generic and modular, and it may be improved to better reduce information transfer loss across modules. As a result, we anticipate that this network will be useful for a wider range of organ volumes or plane segmentation projects, such as blood vascular and brain tumor segmentation. In future studies, we intend to address the model's existing shortcomings to reach a wide range of medical clinical applications. To compare the performance of MPU-Net on medical image segmentation, other evaluation modalities and high-quality models will be applied.